\documentclass[debug,overfull]{epl} 
\usepackage{epsfig}
\usepackage{changebar}
\usepackage{graphicx}
\usepackage{epsfig}
\usepackage{subfigure}
\usepackage{amsmath}
\usepackage{rotating}
\usepackage{amsfonts}
\usepackage{amssymb} 
\newlength {\oldtextheight}
\newlength {\oldheadsep}
\psfigdriver{dvips}

\begin{document}
\title{Mesoscopic  fluctuations and intermittency in  aging  
dynamics} 
\shorttitle{Mesoscopic  statistics and intermittency} 
\author{Paolo Sibani}
\institute{Fysisk Institut, SDU, DK5230 Odense M}
\date{\today}
\maketitle
\begin{abstract}  
Mesoscopic aging systems are  
characterized by large intermittent noise fluctuations.
In a  \emph{record dynamics} scenario [P.~Sibani and  J.~Dall, Europhys. Lett. 64, 2003] these
events, quakes, are   treated  as a Poisson process with  average
$\alpha \ln(1 + t/t_w)$, where  
$t$ is the observation time,   $t_w$ is
 the age and $\alpha$  is a  parameter.  
Assuming for simplicity that   quakes  constitute  the only  source of 
 de-correlation, we present a model for the 
probability density function (PDF)
  of the configuration autocorrelation function. 
 Beside $\alpha$, the model has   the average  quake size  $1/q$
 as a parameter.  The  model autocorrelation PDF   has 
a    Gumbel-like shape, which 
approaches a Gaussian
for  large $t/t_w$  and becomes sharply peaked   in the thermodynamic
limit.  Its average and variance, which are  given analytically, depend
  on $t/t_w$ as a    power-law and a power-law with a logarithmic correction, respectively.
Most    predictions  are in good agreement 
with    data from the literature and with  the  
 simulations of   the   Edwards-Anderson spin glass  carried out as a   test.
\end{abstract}
\vspace{-1cm}
 
\section{Introduction}
After a   rapid quench of an external   parameter, e.g.\ the   temperature,   
 many complex materials  \emph{age}, i.e.\  their properties  
slowly   change with the   \emph{waiting time},   $t_w$, elapsed
from the  quench.  
Ever since  the initial observations  in   polymers~\cite{Struik78}, 
evidence has accumulated   that  
 spin-glasses~\cite{Svedlindh87},  
  type II superconductors~\cite{Nicodemi01}, 
glasses~\cite{Kob00}, and 
soft condensed matter~\cite{Cipelletti05}, among others, 
age in   similar ways, e.g.\ : 
 For observation times   $t \ll t_w$   physical averages are
 nearly constant,  and  autocorrelations and  their  conjugate 
 linear response functions are connected by 
an equilibrium-like fluctuation-dissipation   theorem (FDT). 
  Conversely, for $t \gg t_w $  they visibly drift and the   FDT  is violated.
   As  was recently 
  discovered, the drift  happens in an  
 \emph{intermittent} fashion~\cite{Bissig03,Buisson03},
 i.e.\ through    rare, large, and spatially heterogeneous  
 re-arrangements, which  appear  as    non-Gaussian tails 
 in the probability density function (PDF) of configurational  probes 
 such as  colloidal particle displacement~\cite{Kegel00,Weeks00} and
 correlation~\cite{Cipelletti03a} 
 or voltage noise fluctuations in glasses~\cite{Buisson03a}.
   
As aging phenomena are   similar for a broad class of   interactions, 
we seek a   mesoscopic description, and   assume   that  intermittent events, for short 
  \emph{quakes},    are  the main source of de-correlation in   non-equilibrium
 aging. In the framework of record dynamics~\cite{Sibani03,Sibani05},
  quakes are irreversible and  are triggered by  (energy) fluctuations 
of record magnitude. We show how this   leads to  a description of  the 
 configurational autocorrelation function,   more specifically,   
the  dependence of the  shape  of its PDF on  $t$, $t_w$, the
 temperature $T$ and the system size $N$, which    
resembles  observations for   colloidal gels~\cite{Cipelletti03a}  
   spin-glasses and kinetically constrained models~\cite{Castillo03,Chamon04}. 
  The  model PDF is    closely approximated by the  Gumbel distributions widely 
 used in the literature~\cite{Bramwell00,Chamon04}.  
 The  average and variance are given in close form  as a function of 
 $t/t_w$, $T$ and $N$.   The   average and  the  PDF, standardized to
 zero mean and unit variance, are    in excellent  agreement 
 with   spin-glass simulations. The  agreement is  rather  poor 
 for the variance itself, mainly   because  pseudo-equilibrium fluctuations
 are neglected.  
 
\section{The  configuration auto-correlation  PDF} 
\label{Theory}   
In the model, a  set of $N$ binary variables  defines the system configuration.  Without 
further loss of generality, and with an eye 
 to the simulations of the   E-A spin-glass model~\cite{Edwards75},   
 we refer to these variables  as   spins, and to their changes 
of state  as `flips'.   
 
Configuration changes  are gauged  by  the    
number of spins   with different    orientations  
 at  times $t_w$ and $t_w+t$.  This 
Hamming distance,   $H$,  
is simply related to the autocorrelation $C$ by    
 \begin{equation}
 C(t_w,t) = 1 - 2H(t_w,t_w+t)/N.
 \label{corr_vs_ham}
 \end{equation} 
Initially, we   focus on   the     probability
$P_H(h,t_w+t\mid 0,t_w)$ for  
 $H=h$  at time $t_w+t$, given  $H=0$
at  $t_w$, which we write as the   average
 \begin{equation}
 P_{H}(h,t_w+t\mid 0,t_w) =  \sum_{s=0}^\infty P_{S}(s)P_H(h \mid s) 
 \label{Hamming_unconditional}
\end{equation} 
over the conditional  probability  $P_H(h \mid s)$ for   $H=h$ given 
$s$ flips $(s=0,1,\ldots \infty)$. The weight function 
$P_S(s)$ is   the probability 
for   exactly  $s$    flips  during $[t_w, t_w+t)$.  

Assuming  for simplicity
that   flips  occur at any site   with  
 probability $1/N$ leads    to the   master equation: 
\begin{equation}
P_H(h\mid s+1) = (1-\frac{h-1}{N}) P_H(h-1 \mid s) + \frac{h+1}{N}P_H(h+1\mid s)  
\quad 0\leq h \leq N, 
\label{master_eq}
\end{equation}
with the initial  condition  
\begin{equation}
P_H(h\mid 0) = \delta_{h,0}.
\label{master_init}
\end{equation} 
The equation  has the  formal solution  
\begin{equation}
{\bf P}_H(h \mid s) = {\cal T}^s {\bf P}_H(h\mid 0),
\label{sol}
\end{equation}
where  $\cal T$ is  the  (bi-diagonal) stochastic
matrix    implicitly 
given by Eq.~\ref{master_eq} and  where the vector ${\bf P}_H$ 
has  elements  
 $P_H(0\mid s), P_H(1\mid s) \ldots P_H(N\mid s)$.  
\begin{figure}[t]
\centering
\mbox{
\subfigure[Three model PDF's]{\epsfig{figure=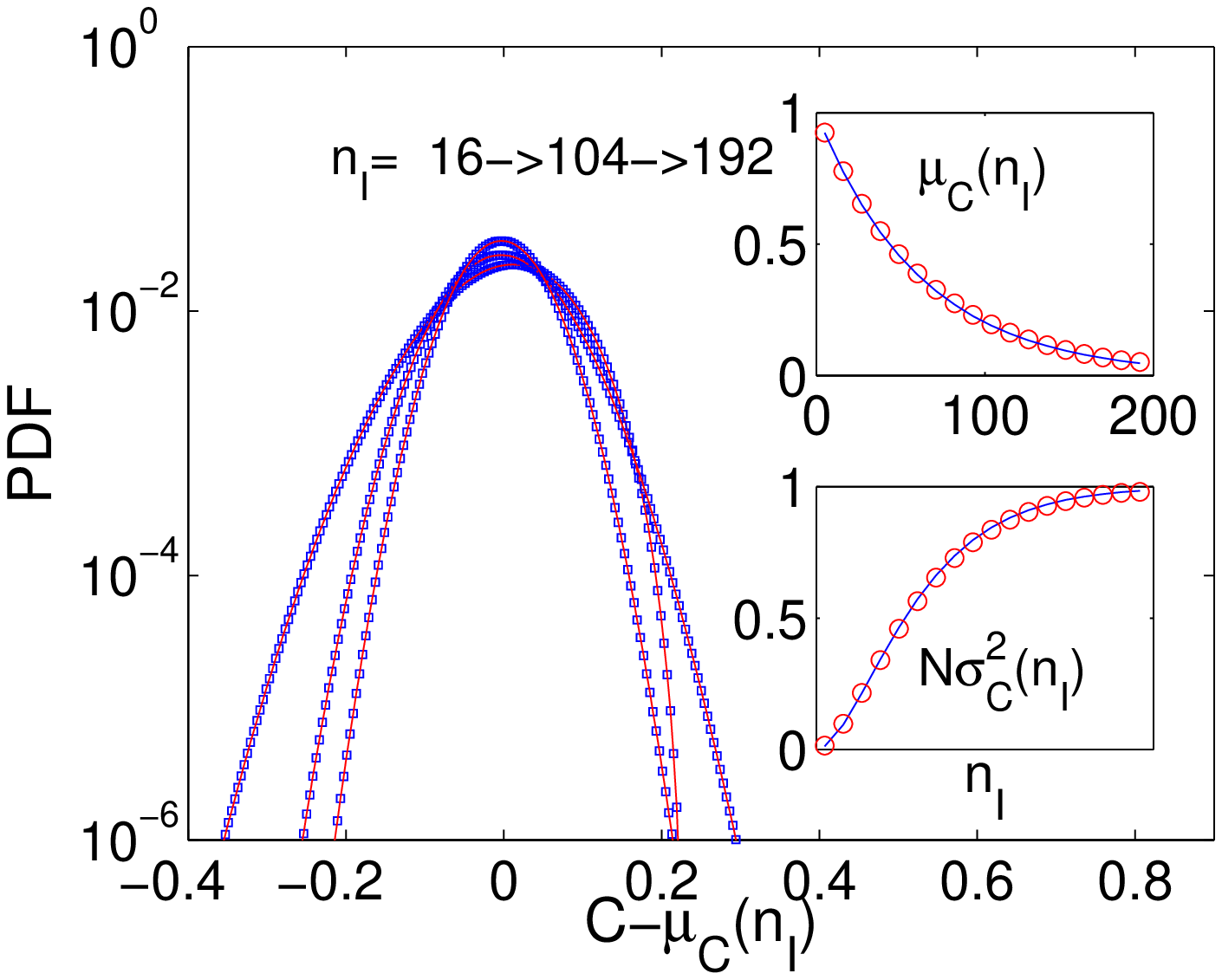,width=.45\textwidth}} \quad 
\subfigure[Comparison with Gumbel densities]{\epsfig{figure=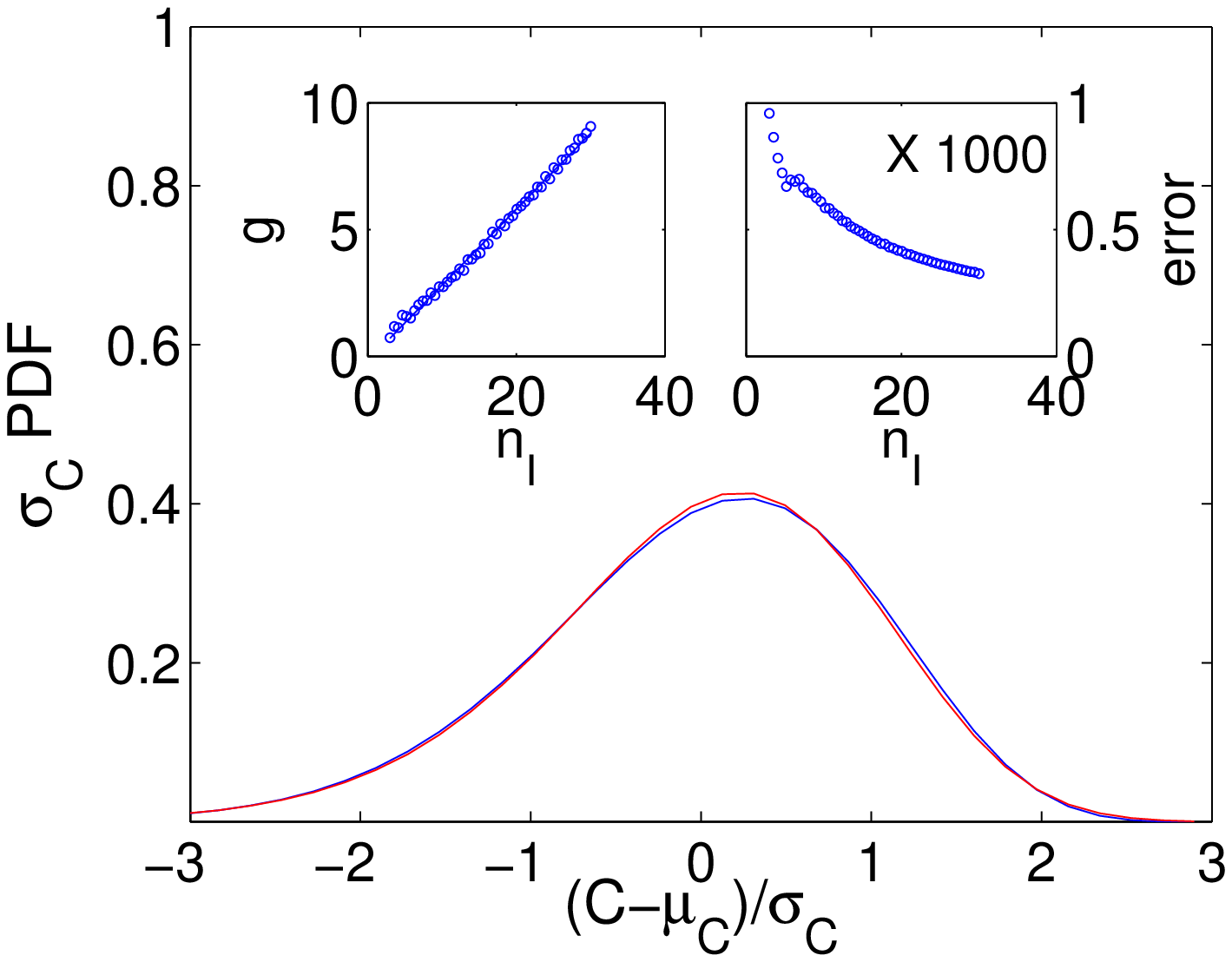,width=.45\textwidth}}} \quad 
\caption{{\em (a)}: Three    PDF's  of the model  auto-correlation function $C$
are shown in the main panel, for different values of   $n_I$. The PDF's are all shifted to zero average.
The  inserts show  the $n_I$ dependence of
the average and the variance, the latter multiplied by $N$.   
 The symbols are from numerical evaluations, and the full lines are  
according to Eqs.~\ref{av_corr_time} and~\ref{var_C_time}.    
{\em (b)} The model PDF  
 for $n_I=16.22$ is shown together with  its  best Gumbel  approximation, 
 with both curves  
  shifted to zero mean and rescaled to  unit variance. 
  The left insert shows the  linear relation
between  the model parameter  $n_I$  
and the $g$ value  for the best Gumbel fit. In the  right panel  the
 approximation error,  (multiplied by $1000$), is plotted versus $n_I$.
  }
\label{PDF_of_correlation}
\end{figure}   
 The $s$ dependence of the conditional average and variance of $H$,
$\mu_H(s)$ and $\sigma^2_H(s)$  can be gleaned from  the  moment generating function 
$ \sum_{h=0}^N P_H(h\mid s)z^h, \quad \mid z \mid \leq 1$.
Omitting the details, one finds  
\begin{equation}
\mu_H(s) =  \frac{N}{2} ( 1 - ( 1-2/N)^s)
\label{av_H}
\end{equation}
and 
 \begin{equation}
\sigma_H^2(s) =    \frac{N}{4} ( 1 - ( 1-4/N)^s) +\frac{N^2}{4}((1-4/N)^s -  (1-2/N)^{2s}).
\label{var_H}
\end{equation}  
Since  $\sigma_H(s) \ll \mu_H(s)$ for  large $N$,    
 the  r.h.s. of Eq.~\ref{Hamming_unconditional} is dominated
in this limit by  the term  with  index $s(h)$ implicitly  given  by $h = \mu_H(s)$. As a consequence,
$P_H$ and $P_S$  acquire very similar shapes when  standardized to zero average and unit variance.
 
 To calculate $P_S(s)$ we need the  probability 
 that  $i$  quakes  occur between $t_w$ and
  $t_w+t$ and the distribution of the number of flips, for short  `size', of 
  each quake. 
 According to  refs.~\cite{Sibani03,Sibani05},   
  $i$   has    a 
  Poisson distribution   with  average 
\begin{equation}
   n_I(t_w,t) =  \alpha(N) \ln(1 + t/t_w). 
\label{average1}
\end{equation}  
The property  $\alpha(N) \propto N$, which removes the $N$ dependence
of the exponent $\lambda$, (see  Eq.~\ref{Lambda}), arises when 
the intermittent signal results  from independent intermittent
processes, stemming e.g.\ from locally thermalized clusters~\cite{Sibani05}. 
The $T$ independence of $\alpha$ reflects
the noise insensitivity  of record dynamics~\cite{Sibani03},  
and  holds within  the   low temperature range  
for  which the  description   applies. 
 
For the     quake size,    
 simulations  of  vortex dynamics~\cite{Oliveira05} yield a near exponential
 distribution. The same  form   is 
  consistent with the  (asymptotically)
exponential distribution of the 
energy released~\cite{Crisanti04,Sibani05} by intermittent 
events. Hence, glossing over the integer nature of the
 sizes,   we   treat them     as  independent stochastic
variables  $X_k$,   $k=1,2\ldots i$, with the     PDF
\begin{equation}
 P_X(x) = q(T) \exp(-q(T)x).
\label{quake_size}
\end{equation} 
A  temperature dependence of the reciprocal  average quake size, $q(T)$, 
is allowed (but not required) by the theory, and is  
directly observable through  the exponent $\lambda$, see Fig.~\ref{mean_and_var_correlation}. 
For typographical clarity this dependence is left understood, together 
with the dependence of $n_I$ on $\ln(1 +t/t_w)$. 
 
Considering  first the conditional  probability for  $S_i$ flips  for a given 
  number $i$ of quakes, we note  that $S_i = \sum_{k=1}^i X_k$    has 
   the  gamma density  
\begin{equation}
 P_{S_i}(x) = q \frac{(qx)^{i-1}}{(i-1)!} \exp(-q x). \quad i>0.
 \label{sum}
\end{equation}  
 Averaging    the above expression  over  the Poisson distribution of $i$, and 
 taking into account  that $\delta(s)\exp(-n_I)$ 
 is the probability of no flips $(i=0)$, 
 one finds 
\begin{equation}
 P_{S}(s) =
  \sum_{i=1}^\infty  P_{S_i}(s) \frac{n_I^i}{i!} \exp(-n_I) + \delta(s)\exp(-n_I).
 \label{sum_2}
\end{equation}  
With  the  variable
$z = (4 q s n_I)^{1/2}$, this is rewritten   as 
\begin{equation}
 P_{S}(s) = 2 n_I a \exp(-qs -n_I)  I_1(z)/z + \delta(s)\exp(-n_I), 
 \label{bessel}
\end{equation}  
where $I_1$ is the  modified Bessel function of 
order one (See e.g. Abramowitz \& Stegun, 9.6.10).

The  $\delta(s)$    term in Eq.~\ref{bessel}
 will be  neglected,  since $n_I$ is large 
 except for  $t \ll t_w$.
With  the term discarded,  
 the  standardized  $P_S(s)$ has no  $T$ dependence. This  
 is seen, in brief,  as follows: 
 Using    $\mu_S=n_I/q $ and $\sigma_S^2=2n_I/q^2$ for the average 
 and variance of $S$, the      standardized PDF,  
$\sigma_S P_S((s-\mu_S)/\sigma_S)$,   has no 
$q$ dependence.  However, as   $q$  
carries  the  model full  $T$ dependence,  
the latter  disappears as well. 
Furthermore,  due to its similarity with $P_S$, 
the standardized $P_H$ is  
 also   independent of $T$, as   
confirmed by  Fig.~\ref{all_correlation}, 

Averaging Eqs.~\ref{av_H} and ~\ref{var_H} over 
$P_S(s)$ , reintroducing  the time dependence of $n_I$ and making  the
approximation $-\ln(1-2/N)\approx 2/N$, one finds
\begin{equation}
\mu_{H}(t_w,t) =  \frac{N}{2} \left(1 - (1 + t/t_w)^{\lambda(T)} \right), 
\label{av_H_time}
\end{equation} 
where 
\begin{equation}
 \lambda(T) = -2 \frac{\alpha(N)}{N} \frac{1}{q(T)}.
 \label{Lambda}
 \end{equation} 
The average (macroscopic) form of the   correlation function $C$,  
 is obtained from Eqs.~\ref{av_H_time} and~\ref{corr_vs_ham} as 
\begin{equation}
\mu_{C}(t_w,t) =    (1 + t/t_w)^{\lambda(T)}. 
\label{av_corr_time}
\end{equation}
Similar steps lead from  Eq.~\ref{var_H} to
 \begin{equation}
N \sigma^2_{C}(t_w,t) = 1- (1 + t/t_w)^{2\lambda(T)}\left(1 -2\lambda(T) \ln(1+t/t_w)   \right).  
\label{var_C_time}
\end{equation}   
Panel \emph{(a)} of Fig.~\ref{PDF_of_correlation} shows, for three different values of 
$n_I=\alpha(N) \ln(1+t/t_w)$, 
  the  model  PDF given by    Eqs.~\ref{bessel},~\ref{sol} and~\ref{Hamming_unconditional}.  
 The two inserts  show the 
$n_I$ dependence of   $\mu_{C}$ and  $N\sigma^2_{C}$ 
from a numerical evaluation of the model equations (circles) and from 
Eqs.~\ref{av_corr_time} and \ref{var_C_time}
with the $n_I(t,t_w)$ dependence reintroduced (lines). 

In standard form (see e.g. ref~\cite{Chamon04}), 
the  one-parameter family of Gumbel densities 
 is  given by $\Phi_g(y) = \frac{\mid b \mid g ^g}{\Gamma (g)}\exp(b(y-y_0)-e^{b(y-y_0)})$,
with  $b=\surd(\Psi'(g))$ and $b y_0 = \ln(g) -\Psi(g)$, where 
 $\Psi$ denotes  the digamma function,  $\Psi'$  its derivative
and $g$ is a real number. Gumbel densities 
empirically  describe   fluctuations in 
 complex systems~\cite{Chamon04,Bramwell00}.
    Fig.~\ref{PDF_of_correlation},\emph{(b)} shows that, except for 
$n_I <1$,  our model   
PDF is closely  approximated by the  Gumbel  PDF whose  $g$ value  minimizes
the $L_1$ distance between the two. 
The left insert of the figure shows that  this  optimal  $g$ value 
is linearly related to   $n_I$ as   
$ g = 0.300 n_I -0.185$.  
 
\section{Comparison with  simulation data}  
The   (average)   autocorrelation function $\mu_{C}$  
for   spin-glasses is well investigated~\cite{Kisker96,Picco01,Berthier02,Chamon04}. 
For $t>t_w$,    $\mu_{C}$ is   nearly  a function of $t/t_w$,  and   can be 
 fitted by   a power-law  with a temperature dependent exponent.  E.g.\ Picco et al.~\cite{Picco01} 
  found an excellent scaling  using the variable $\ln(t_w+t) - \ln(t_w)$.
\begin{figure}[t]
\centering
\mbox{
\subfigure[mean of autocorrelation]{\epsfig{figure=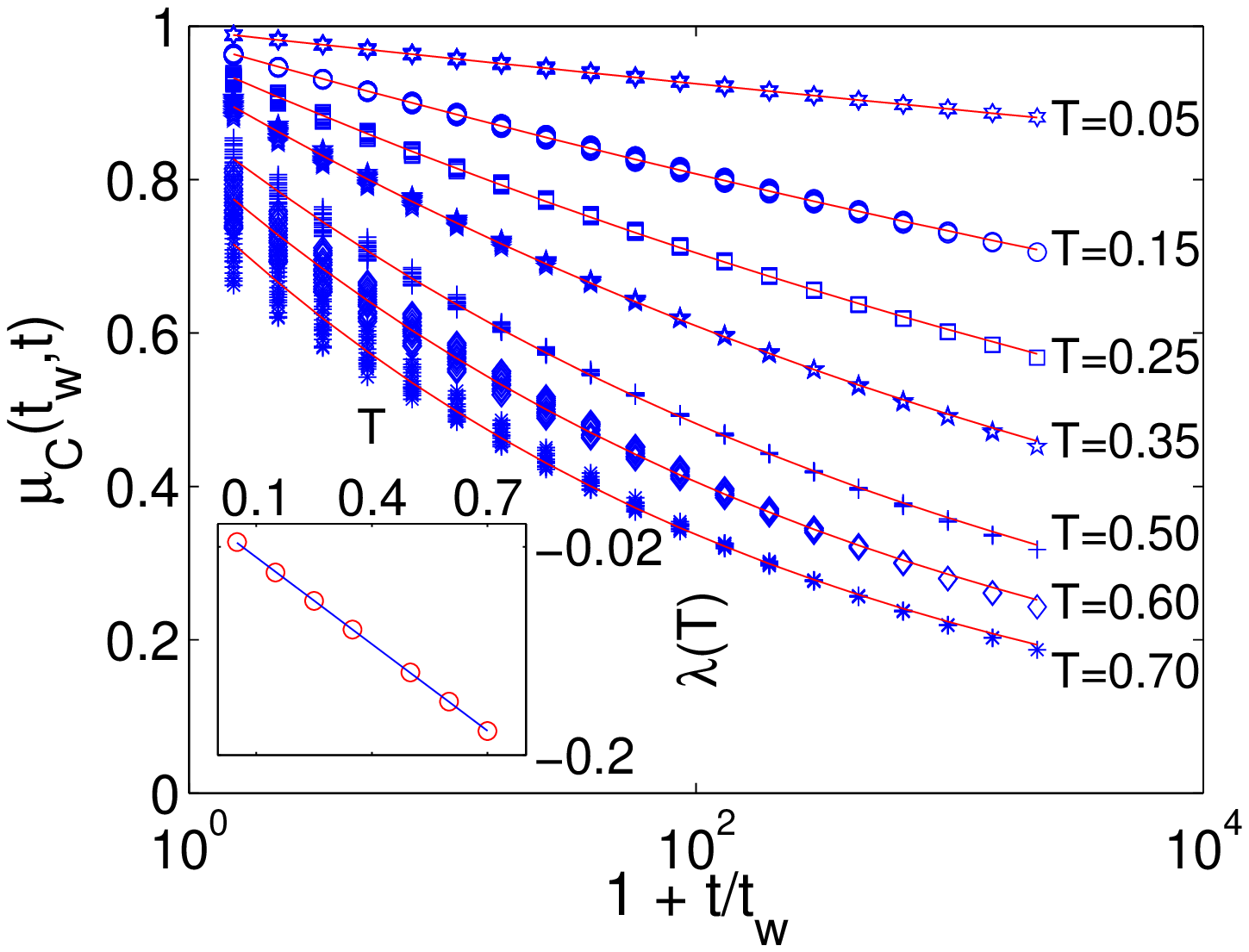,width=.47\textwidth}} \quad 
\subfigure[variance of autocorrelation]{\epsfig{figure=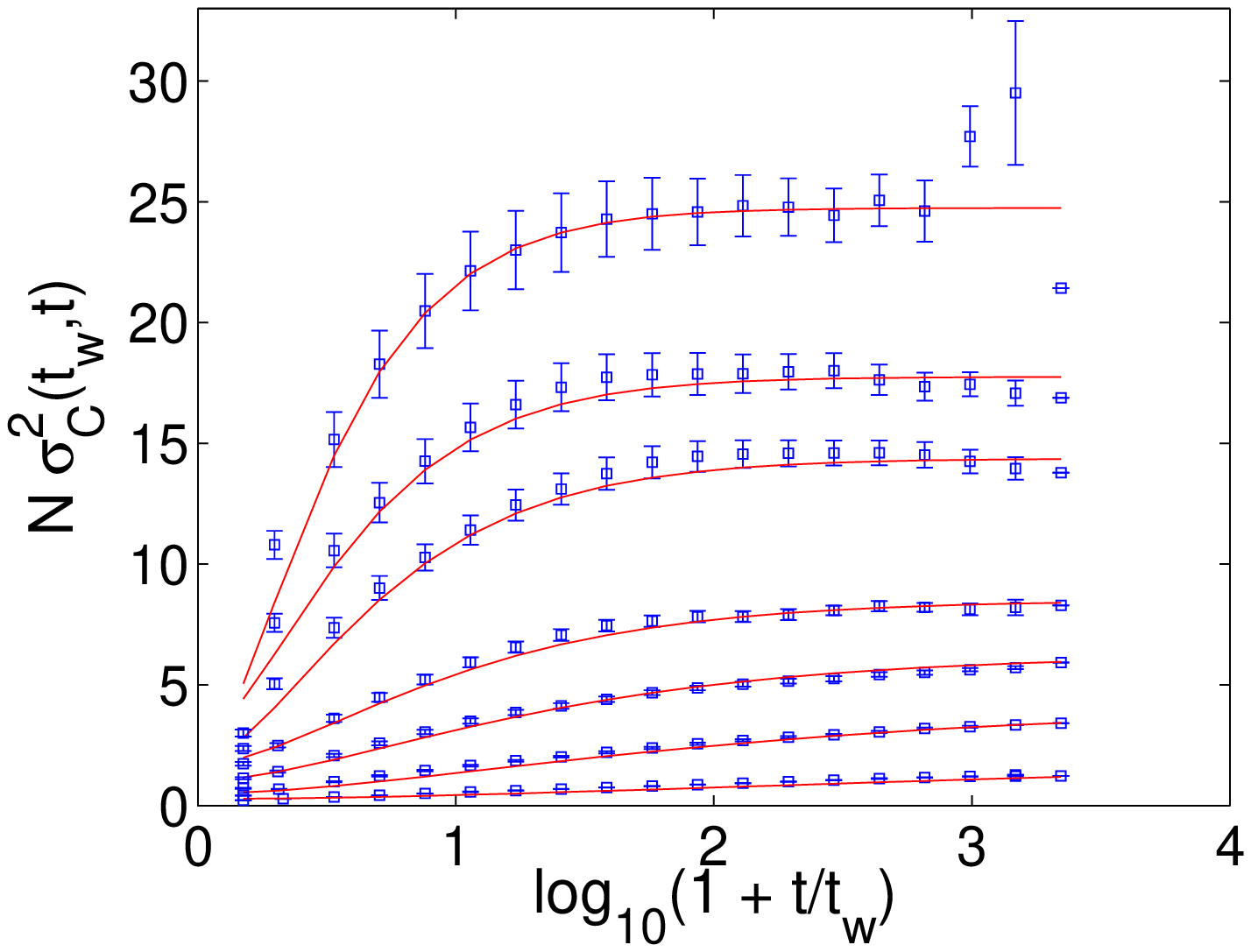,width=.47\textwidth}}} \quad 
\caption{{\em (a)}: The    mean of the autocorrelation function
in an   E-A spin-glass of size $N=16^3$ for different values of $t_w$ and $t$ is shown 
by symbols  versus  $\ln(1+t/t_w)$, for  the indicated  temperatures.
 The full lines are power-law fits according to Eq.~\ref{av_corr_time}.
 The insert shows the   temperature
dependence of  the exponent $\lambda(T)$.    
{\em (b)} For the  same temperatures, the estimated variances of the  autocorrelation function
multiplied by 
versus $\ln(1+t/t_w)$ together with 1$\sigma$ error-bars. The full lines are  fitted  to  
a shifted and rescaled  form of  Eq.~\ref{var_C_time}, as detailed in the   text. }
\label{mean_and_var_correlation}
\end{figure} 
The   autocorrelation   PDF for the E-A spin glass model 
nearly follows   $t/t_w$ scaling according to 
Castillo et al.~\cite{Castillo03} 
Chamon et al.~\cite{Chamon04}  also consider  a kinetically constrained model with trivial statics. 
In both cases,  the autocorrelation PDF, shifted to zero mean and
rescaled to unit variance,  is empirically  
 fitted to  time evolving Gumbel distributions, which are
 numerically equivalent to our model results (see  Fig.~\ref{PDF_of_correlation}). 

For  a more detailed comparison, we  simulated  the  E-A  spin-glass~\cite{Edwards75}
 on a cubic lattice with $N=16^3$   using 
an      event driven simulation technique~\cite{Dall01}, 
whose  `intrinsic'   time unit   
corresponds,  for large systems, to 
one  Monte Carlo sweep.
 The  data are  sampled at 
$20$     time points,  which are  separated by a multiplicative factor of $1.5$, with  start   
  at  $t=100$ and  end  at $t\approx 2.2 \times 10^5$. 
Among these  points, any ordered pair   can be chosen  for  $t_w$ and $t_w+t$.
For each set of physical parameters,  $1000$  runs are performed  with independent 
 noise and     couplings realizations, producing e.g.\  $1000$ data points  for 
$t/t_w=2216$, and $20000$ points  for $t/t_w=1.5$.  
 
For a range  of temperatures, the average spin-glass autocorrelation 
is plotted versus $t/t_w$    (symbols). 
 Deviations    from  $t/t_w$ scaling  appear  for approx.    
$t/t_w < 4 $ and $T>0.5$, as seen  
   in  the left panel of Fig.~\ref{mean_and_var_correlation} 
from  the poor  data collapse. Away from this parameter region,   the
 data  are well described  by 
 Eq.\ref{av_corr_time} (full line). 
The  $T$ dependence of the exponent is  shown in the insert (circles), 
where  the fit $\lambda(T) =  -0.25 T/T_g$ is also shown (full line).  $T_g=0.97$ is the critical temperature
of the model~\cite{Marinari98}. The same linear form,   
and with  a similar  slope coefficient,  is found numerically  in  Kisker et al.~\cite{Kisker96}.
They, however,  introduce 
a small $t_w$ dependence of $\lambda$, which is   beyond  the present model.
  
Summarizing, for low $T$  and  
not too small  $t/t_w$,  the model is able  
to describe the  time  dependences of the average autocorrelation   with no 
free parameters. Furthermore,   
Eq.~\ref{Lambda}  links    $\lambda(T)$ to a linear 
temperature increase of the  average quake size.

To  improve the   statistics of the variance and PDF 
data in the $t/t_w$ scaling region 
where  a  comparison with the model is most interesting,  
we estimate the   variance and its error-bar    
as        the mean value  
and standard deviation over the set 
of   variances for all    pairs  $t_w,t$ having   the same 
ratio $t/t_w$. 
Similarly, the empirical frequencies   of  the   $C$ values 
are calculated based on  all  data  with the same  $t/t_w$.  

\begin{figure}[t]
 \vspace{-.7cm}
\begin{center}
 \epsfig{figure=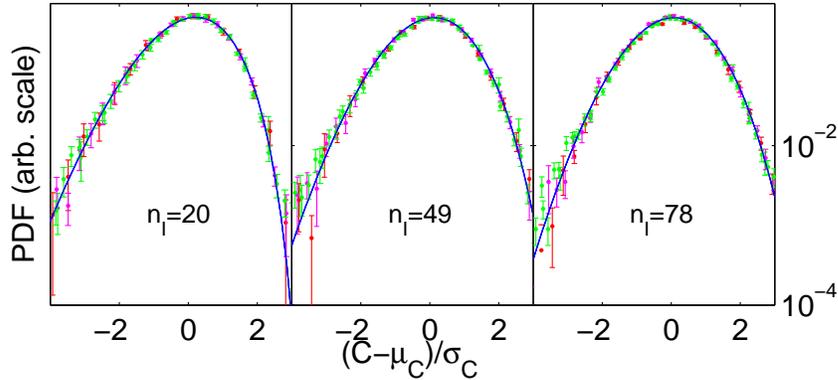,width=.9\textwidth} 
 \end{center}
 \vspace{-0.95cm}
\caption{ From the   left to the right panel,
the   $n_I$  values indicated correspond to   
 $t/t_w = 2.3, 7.6$ and $25.6$. 
Within each  panel,  the scaled and shifted  autocorrelation  PDF 
is shown  for the model (full line) together with four   sets of  simulation data
for  temperatures $T=0.15, 0.25, 0.35$ and $0.5$. 
}  
\label{all_correlation}
\end{figure}

 The value of $N$   stands in the model for an  (unknown) number of thermally active spins,
 and appears   in the autocorrelation   variance, which  
 vanishes linearly with  $1/N$. This leads   
 to    an undetermined
 $T$ dependent    scale factor,  $f_1$, between model and  simulation variance.
Secondly, and more importantly,  the data  cannot  be fitted without
a second, ad hoc, offset parameter $f_2$, likely 
because the  de-correlating effect of the pseudo-equilibrium
fluctuations is altogether  neglected. Hence, with respect to the variance,
the theory  only provides  a qualitative 
description, which 
  is captured   by the  empirical formula    
$N \sigma^2_{emp.}(t_w,t) =   f_1 (N \sigma^2_{C}(t_w,t) + f_2)$. 
For completeness, the latter 
is  plotted (lines)   in the   right panel of Fig.~\ref{mean_and_var_correlation}
together with the  simulation data   with error bars. The parameter 
$f_1$ increases with $T$  within the range $1-20$,
and  $f_2$ remains  close to  $1/10$,   

By contrast, an excellent agreement  between 
 predictions and data   for the standardized PDF is 
 achieved   by a simple adjustement 
of  the   vertical scale of   the latter. This scale,    which is
undetermined  from the outset,
is fitted to   the properly normalized model PDF.  The   centering  and rescaling
are done using the data average and standard deviation. The results  are plotted with   $1\sigma$ error-bars in 
Fig.~\ref{all_correlation}.   
The   three panels  of the figure correspond, from left to right,  to  
$t/t_w=2.3$,   $7.6$ and $25.6$.  In each panel, the data 
shown  are for   $T=0.15, 0.25, 0.35$ and $0.5$. Their collapse 
confirms the    anticipated $T$ independence of the standardized  
autocorrelation PDF. 
  The model predictions (full lines) 
 contain one   parameter, $\alpha(N)$, whence it is  possible to 
determine one value of $n_I$ by data  fitting. This was done
 for  $(t/t_w=25.6)$---in the rightmost panel---yielding   $n_I=78$ .  Eq.~\ref{average1} then gives 
$\alpha=24$, whence  $n_I=20$ and $49$ for $t_w=2.3 $ and $7.6$ respectively.   

\section{Conclusion}
Based on   the   record  dynamics description of intermittency~\cite{Sibani03,Sibani05}      
the  model develops the    aging  properties 
 of the   configuration autocorrelation
after a deep quench. Its predictions  for the average autocorrelation and the 
standardized PDF are 
accurate at low temperatures and for $t > t_w$. 
Together   with allied  efforts~\cite{Dall03,Sibani04a,Boettcher05,Oliveira05},  
 the present  results   support the view  that  record-sized fluctuations
are important for  aging in metastable glassy systems. 
 \section{Acknowledgments} Support  from the Danish Natural Sciences Research Council
is gratefully acknowledged.   
\bibliographystyle{unsrt}
\bibliography{SD-meld}
\end{document}